\documentclass[10pt,conference]{IEEEtran}
\pdfoutput=1
\IEEEoverridecommandlockouts

\usepackage{cite}
\usepackage{amsmath,amssymb,amsfonts}
\usepackage{graphicx}
\usepackage{textcomp}
\usepackage{xcolor}
\usepackage{enumitem}
\usepackage{booktabs}
\usepackage{algpseudocode}
\usepackage{colortbl}
\usepackage{listings}
\usepackage{pifont}        
\usepackage{fontawesome5}
\usepackage{seqsplit}
\usepackage{stfloats}
\usepackage{multirow}
\usepackage{url}
\usepackage[ruled,linesnumbered]{algorithm2e}
\usepackage[most]{tcolorbox}
\usepackage{balance}

\definecolor{cypherblue}{RGB}{0, 0, 255}
\definecolor{cyphergray}{RGB}{128, 128, 128}
\definecolor{bglightgray}{RGB}{248, 248, 248}
\definecolor{bugred}{RGB}{220, 38, 38}
\definecolor{boxgray}{RGB}{245,245,245} 
\definecolor{titlegray}{RGB}{140,140,140}

\newtcolorbox{promptbox}{
    colback=boxgray,
    colframe=black,
    boxrule=0.8pt,
    arc=2pt,
    outer arc=2pt,
    title=\textbf{\textcolor{white}{Prompt Template}},
    coltitle=white,
    colbacktitle=titlegray,
    fonttitle=\bfseries,
    left=6pt,
    right=6pt,
    top=6pt,
    bottom=6pt
}

\newtcolorbox{rqanswer}{
    enhanced,
    breakable,
    nobeforeafter, 
    colback=gray!6,
    colframe=black!70,
    boxrule=0.8pt,
    arc=2.5mm,
    outer arc=2.5mm,
    left=3mm,
    right=3mm,
    top=2.2mm,
    bottom=2.2mm,
    width=\linewidth,
    boxsep=0pt,
}

\lstdefinestyle{cypherstyle}{
    backgroundcolor=\color{black!5},      
    basicstyle=\ttfamily\small\color{black}, 
    keywordstyle=\bfseries\color{black},  
    commentstyle=\itshape\color{gray},    
    numbers=left,
    numberstyle=\tiny\color{black},
    stepnumber=1,
    numbersep=8pt,
    frame=single,               
    rulecolor=\color{black},    
    breaklines=true,
    breakatwhitespace=true,
    showstringspaces=false,
    escapeinside={(*@}{@*)},
    xleftmargin=8pt,       
    xrightmargin=5pt
}

\begin{document}

\title{Graph-Aware Fuzzing for Graph Database Management Systems
\author{
\IEEEauthorblockN{
Yu Li\textsuperscript{1}, Qiang Hu\textsuperscript{1}, Yao Zhang\textsuperscript{1}, Junjie Wang\textsuperscript{1}, Hao Liu\textsuperscript{1}, Rui Wang\textsuperscript{2}, Yongqiang Lyu\textsuperscript{1}
}
\IEEEauthorblockA{
\textsuperscript{1}\textit{Tianjin University}, Tianjin, China \\
\textsuperscript{2}\textit{Capital Normal University}, Beijing, China \\
\{liyu2025, qianghu, zzyy, junjie.wang\}@tju.edu.cn
}
}
}

\maketitle

\begin{abstract}
Graph Database Management Systems (GDBMSs) are essential infrastructure for managing interconnected data. Existing GDBMS testing methods primarily rely on differential and metamorphic testing. The result consistency oracles of these methods constrain inputs to queries that are comparable across engines or transformations, leaving single engine runtime failures, such as crashes and memory errors, insufficiently explored. Developing dedicated fuzzers for GDBMSs faces two key challenges: (1) generating valid and structurally diverse queries under complex graph constraints, and (2) guiding exploration to capture topology dependent execution behavior.

To address these challenges, we propose \textsc{GRAF}, a black box fuzzing framework for GDBMS query engines. First, \textsc{GRAF} introduces graph context aware query generation based on cascading dependency resolution. It instantiates parameterized Cypher skeletons generated by a Large Language Model (LLM) by jointly resolving labels, relationship types, properties, values, and variable scopes against the active graph state. This process produces structurally diverse queries while eliminating syntactic and semantic violations. Second, \textsc{GRAF} applies five graph specific mutation operators guided by execution state feedback, including execution time, result size, and system status. This feedback steers exploration away from unproductive queries and expensive traversals, while prioritizing local mutations around abnormal executions. 

We evaluated \textsc{GRAF} against three existing approaches on six widely used GDBMSs. \textsc{GRAF} consistently improves line coverage by 31.6\% to 41.1\% over the strongest baseline on each target. In 12 hour fuzzing, it triggered 25 unique bugs, compared to six from all baselines combined. Overall, \textsc{GRAF} discovered 34 previously unknown bugs, with 32 confirmed by developers and 23 assigned CVEs.
\end{abstract}


\section{Introduction}
\label{sec:introduction}

Graph Database Management Systems (GDBMSs), such as Neo4j~\cite{neo4j2010}, Memgraph~\cite{memgraph2020}, and RedisGraph~\cite{redisgraph2018}, have become essential infrastructure for managing highly connected data. Unlike relational database management systems~\cite{rdbms,rdbms1} that rely on rigid schemas and table based joins, GDBMSs utilize the labeled property graph model to represent data naturally as nodes, relationships, and properties. This architecture simplifies the execution of complex multi hop queries, driving their adoption in domains like financial fraud detection, social network analysis, and real time recommendation~\cite{recommender}. As these applications depend on continuous availability, ensuring the reliability of the underlying database engine is paramount. Defects within a graph query engine can lead to system failures, including runtime crashes, memory corruption, and resource exhaustion.

General DBMS fuzzers have proved effective for testing relational database systems. Tools such as SQLsmith~\cite{sqlsmith2018} and SQLancer~\cite{TLP,PQS,Norec} generate SQL queries based on abstract syntax tree models or formal semantics, successfully exposing bugs in mature systems such as PostgreSQL~\cite{postgresql} and SQLite~\cite{sqlite}. More recent fuzzers, including SQUIRREL~\cite{SQUIRREL} and Griffin~\cite{Griffin}, further improve input validity by preserving syntax or exploiting database metadata. These successes show that fuzzing is an effective way to exercise deep database execution logic and expose runtime failures.

\begin{figure}[t]
    \centering
    \includegraphics[width=\linewidth]{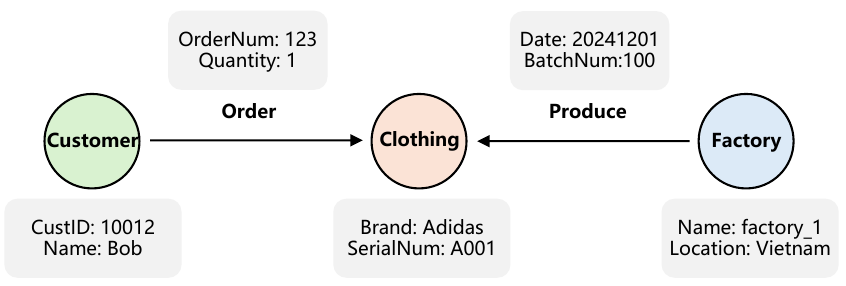}
    \caption{Core elements of a labeled property graph: labeled nodes, directed relationships, and typed properties.}
    \label{fig:lpg}
\end{figure}

However, dedicated fuzzing techniques for GDBMSs remain underexplored. Existing approaches mainly rely on differential testing~\cite{GDsmith,GQS,Grand} and metamorphic testing~\cite{GRev,GraphGenie,GDBMeter,QuDi,Gamera}. Since these techniques are built around result consistency oracles, they typically require generated queries to be comparable across different engines or semantically related transformations. This requirement narrows the explored input space toward well defined and result producing queries, leaving engine specific runtime failures, such as crashes and memory errors, insufficiently explored. Designing effective fuzzing techniques for GDBMSs remains difficult due to two key challenges:

\textbf{Challenge 1: Generating complex and valid graph queries.} A valid Cypher query~\cite{cypher} must satisfy interdependent syntactic and semantic constraints, including variable scoping across multi clause pipelines, expressions with compatible types, and relationship patterns that match the labels and relationship types in the underlying graph. These constraints are interdependent: once a source label is selected, the valid relationship types, destination labels, and available properties are all restricted by the graph context. As illustrated in Figure~\ref{fig:lpg}, selecting a \texttt{Customer} node constrains the query to valid outgoing relationships such as \texttt{Order}, which further determines the reachable destination label \texttt{Clothing} and the properties available for subsequent predicates.
As demonstrated in Figure~\ref{fig:motivating_semantic}, generic byte level fuzzers and grammar-based generators struggle to preserve these constraints. Byte mutation frequently corrupts strict Cypher syntax, while blind structural mutation generates semantically invalid patterns that violate graph schemas. Consequently, these invalid inputs are rejected by the frontend parser, failing to reach the optimization and execution logic of the target GDBMS~\cite{SQUIRREL}. To bypass the frontend, grammar-based generators~\cite{sqlsmith2018,Dinkel} produce valid but structurally simple queries, failing to thoroughly exercise deep engine logic.

\textbf{Challenge 2: Guiding exploration under topology dependent execution.} Existing GDBMS testing methods~\cite{DQP,Gamera,GDBMeter,GDsmith,GQS,Grand,GraphGenie,GRev} mainly rely on differential or metamorphic testing and typically lack feedback driven exploration. Meanwhile, traditional coverage feedback is insufficient for graph query fuzzing because it captures control flow execution but not topology dependent data flow behavior. Graph query execution is highly sensitive to hubs, cycles, dense subgraphs, traversal depth, and predicate selectivity. For example, on a dense 15 node graph in FalkorDB, increasing a variable length traversal bound from \texttt{[:E1..5]} to \texttt{[:E1..8]} raises execution time from 146~ms to 46.81~s, although the executed control flow paths remain similar. Conversely, highly selective predicates may lead to empty results and little meaningful backend stress. Therefore, a graph fuzzer needs lightweight execution state feedback beyond coverage alone.

Our key insight is that query structure and graph specific bindings should be separated: diverse query skeletons can be generated, while labels, relationships, properties, values, and scopes are resolved at runtime. In addition, execution time, result size, and execution status provide lightweight signals for distinguishing empty executions, expensive traversals, and abnormal failures. Based on these insights, we present \textsc{GRAF}, a black box, feedback driven fuzzer for GDBMS query engines.

\textbf{Graph Context Aware Query Generation.}
\textsc{GRAF} separates abstract query structure from concrete graph data. It utilizes a Large Language Model to synthesize parameterized query skeletons with diverse clause combinations and traversal patterns. During fuzzing, \textsc{GRAF} binds these skeletons to the active graph context using cascading dependency resolution. By resolving labels, relationships, and properties against the actual graph state, this design preserves structural diversity while eliminating syntactically and semantically invalid queries before execution.

\textbf{Execution State Guided Query Mutation.}
\textsc{GRAF} guides query mutation using a dynamic feedback vector containing execution time, result size, and system status. The framework adapts to these signals by relaxing predicates for queries that return empty results, restricting depth expansion for queries nearing timeout limits, and prioritizing local mutations upon encountering abnormal behaviors. This continuous loop operates entirely as a black box using externally visible outcomes. It effectively steers exploration toward deep engine logic without relying on traditional code coverage.

\begin{figure}[htbp]
    \centering
    \begin{lstlisting}[
        style=cypherstyle,
        numberstyle=\tiny,
        basicstyle=\scriptsize\ttfamily,
        escapeinside={(*@}{@*)} 
    ]
// 1. General mutation corrupts Cypher structure
MATCH (*@\textcolor{red}{\textbf{c:Customer}}@*)-[:Order]->(*@\textcolor{red}{\textbf{p:Clothing}}@*) RETURN p.Brand;
(*@\textbf{>> Syntax Error: }@*)
(*@\textbf{>> Node patterns must be enclosed in parentheses `()`}@*)

// 2. Mutation violates graph schema constraints
MATCH (c:Customer)-[:(*@\textcolor{red}{\textbf{Produce}}@*)]->(p:Clothing) WHERE (*@\textcolor{red}{\textbf{c.Location}}@*) = 'Vietnam' RETURN p;
(*@\textbf{>> Semantic Error: }@*)
(*@\textbf{>> `Produce` and `Location` undefined for `Customer`}@*)
    \end{lstlisting}
    \caption{General mutation techniques frequently generate queries with syntax errors or semantic schema violations. These invalid inputs are rejected by the frontend before reaching the core execution engine.}
    \label{fig:motivating_semantic}
\end{figure}

We evaluate \textsc{GRAF} on six GDBMSs: Memgraph, FalkorDB, Neo4j, KuzuDB, NebulaGraph, and RedisGraph. \textsc{GRAF} consistently improves source line coverage by 31.6\% to 41.1\% over the strongest baseline on each target. Furthermore, it has discovered 34 previously unknown bugs. At the time of writing, 32 of these bugs are confirmed, with 23 assigned CVE identifiers.

This paper makes the following contributions:
\begin{itemize}[leftmargin=*]
    \item We introduce \textsc{GRAF}, a fuzzing framework driven by execution feedback to test diverse graph database engines. \textsc{GRAF} operates entirely as a black box and does not rely on traditional code coverage as a feedback signal.
    \item We design a query generation approach decoupling abstract structure from concrete graph data. It synthesizes diverse query skeletons using a LLM and employs cascading dependency resolution for instantiation.
    \item We propose a mutation strategy guided by execution state feedback. By tracking execution time, result size, and system status, \textsc{GRAF} adapts its mutation operators to avoid unproductive queries and uncontrolled path explosions, effectively steering the exploration toward deep engine logic.
    \item We evaluate \textsc{GRAF} on six popular GDBMSs against recent baselines. \textsc{GRAF} improves line coverage and has discovered 34 previously unknown bugs, with 32 confirmed and 23 assigned CVEs.
\end{itemize}

\section{Background}
\subsection{Cypher Query Language}
Unlike SQL used in relational databases, graph databases rely on specialized query languages to handle complex topological logic. Cypher is the most popular and widely adopted declarative graph query language~\cite{openCyper}. It follows a pattern matching paradigm, where queries describe graph structures using nodes, relationships, and properties. A Cypher query is composed of multiple clauses (e.g., \texttt{MATCH}, \texttt{WHERE}, \texttt{WITH}, \texttt{RETURN}) that execute in a pipeline. Beyond basic pattern matching, Cypher supports advanced constructs such as variable length traversals, aggregation, and subqueries, enabling expressive graph operations.

Cypher queries must satisfy multiple layers of validation before execution: (1) syntactic correctness defined by the grammar; (2) schema consistency for labels, relationships, and properties; (3) type correctness for expressions; and (4) scope correctness, where variables must be explicitly propagated across clauses (e.g., via \texttt{WITH}). These constraints are strictly enforced by the frontend, causing invalid queries to be rejected early. As a result, naive query generation approaches produce a large number of non-executable queries, limiting their ability to reach the execution engine.

\begin{figure}[t]
    \centering
    \begin{lstlisting}[
        style=cypherstyle,
        numberstyle=\tiny,
        basicstyle=\scriptsize\ttfamily,
        escapeinside={(*@}{@*)} % 开启逃逸符，允许在代码块中使用 LaTeX 语法
    ]
// Create a dense directed graph with 15 nodes
UNWIND range(1, 15) AS i CREATE (:Node {id: i});
MATCH (a:Node), (b:Node) WHERE a.id <> b.id CREATE (a)-[:E]->(b);

// 1. A 5-hop traversal executes quickly
MATCH p=(start:Node {id: 1})-[:E*1..(*@\textbf{5}@*)]->(end:Node) RETURN count(p);
(*@\textbf{>> Execution Time: 146.00 ms}@*)

// 2. An 8-hop traversal triggers state explosion
MATCH p=(start:Node {id: 1})-[:E*1..(*@\textbf{8}@*)]->(end:Node) RETURN count(p);
(*@\textbf{>> Execution Time: 46.81 s} \textcolor{red}{\textbf{($\approx$ 320$\times$ slower)}}@*)
    \end{lstlisting}
    \caption{Increasing a traversal bound from five to eight on a 15-node dense graph increases execution time by more than 320$\times$.}
    \label{fig:motivating_explosion}
\end{figure}

\subsection{Graph Topology and Execution Behavior}

Unlike the relatively uniform schemas in relational databases, real world graph data is highly irregular. Applications such as social networks, financial transaction networks, and supply chains naturally form diverse topologies, often exhibiting three key patterns: (1) \textit{Hub nodes}, where a few central nodes have extremely high connectivity~\cite{hubs}; (2) \textit{Dense subgraphs}, with tightly connected nodes generating an immense number of internal paths~\cite{subgraph}; and (3) \textit{Cyclic dependencies}, forming loops that allow repeated reachability.

These structural patterns directly shape query execution in GDBMSs. Native graph operations, such as variable length traversals and pattern matching, must dynamically expand paths and materialize intermediate results~\cite{materialize}. The topology dictates both branching and search depth: traversals through hubs trigger sudden expansions; dense subgraphs produce a combinatorial number of candidate paths even with strict bounds; and cycles cause repeated work and backtracking, making execution highly unpredictable and resource intensive~\cite{resource-intensive}.

Consequently, query execution is extremely sensitive to the underlying topology. Even small parameter changes (e.g., increasing path bounds) can trigger an exponential explosion of intermediate states, quickly exhausting CPU and memory and causing fatal out of memory (OOM) crashes. As shown in Figure~\ref{fig:motivating_explosion}, evaluating a 5-hop traversal on a 15-node fully connected graph takes only 146 milliseconds, whereas increasing the bound to 8 hops raises execution time to 46.81 seconds. This slight change forces the engine to materialize a massive number of cyclic paths.

Importantly, traditional metrics such as code coverage only track control flow transitions and fail to capture these critical data flow effects. Both the 5-hop and 8-hop queries traverse the same internal loops without generating new control flow edges, so conventional fuzzers consider the resulting resource spike as uninteresting. This mismatch between actual execution cost and control flow signals demonstrates why existing testing approaches fall short and underscores the need for an execution state feedback mechanism to guide fuzzing effectively~\cite{feedback}.

\section{Design of \textsc{GRAF}}
\label{sec:design}

\subsection{Overview}
\begin{figure*}[t]
    \centering
    \includegraphics[width=\linewidth]{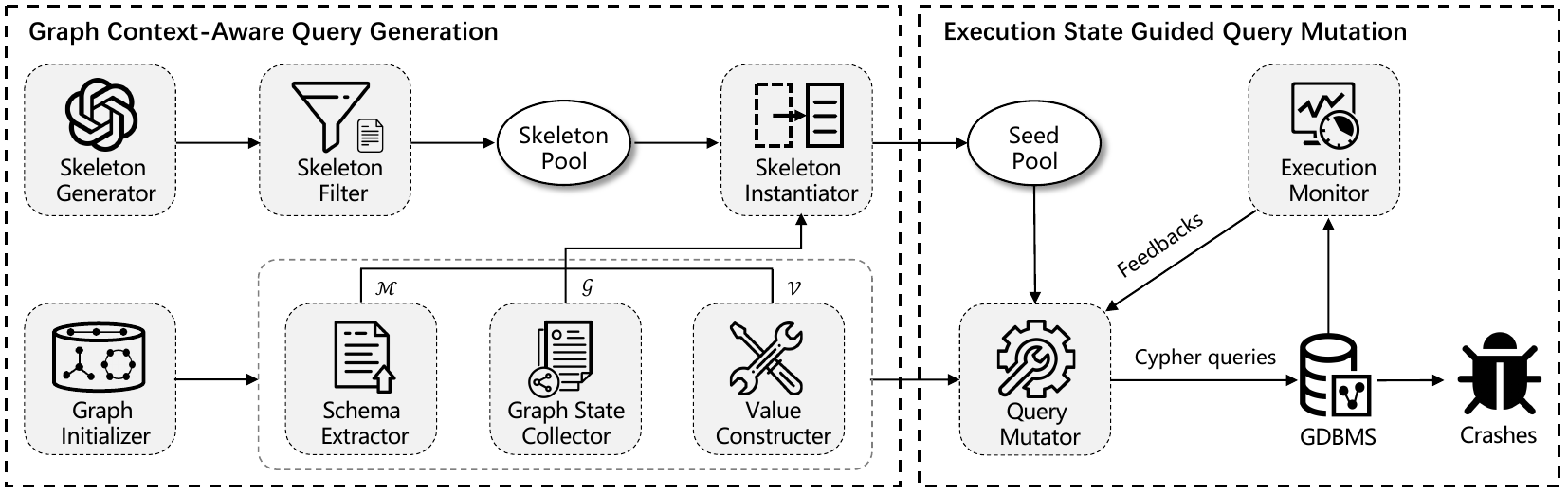}
    \caption{Architecture of \textsc{GRAF}. The left component generates valid and diverse query seeds by combining skeleton filtering with graph context instantiation. The right component performs feedback guided mutation to explore deep GDBMS execution behavior.}
    \label{fig:overview}
\end{figure*}

\textsc{GRAF} consists of two core components in Figure~\ref{fig:overview}: graph context aware query generation and execution state guided query mutation.

The first component constructs executable and structurally diverse query seeds. A skeleton generator produces parameterized Cypher skeletons equipped with standardized variables and typed placeholders. A filter parses each skeleton, checks placeholder and scope constraints, removes redundant structures, and outputs a reusable skeleton pool. In parallel, a graph initializer builds a representative target graph, after which \textsc{GRAF} extracts a graph context $\mathcal{C}=(\mathcal{M},\mathcal{G},\mathcal{V})$. Here, $\mathcal{M}$ records labels, relationship triples, properties, and data types; $\mathcal{G}$ summarizes graph scale and topology; and $\mathcal{V}$ provides observed, boundary, and anomalous values. The instantiator uses this context to bind placeholders and produce executable query seeds for the seed pool.

The second component explores the target database using feedback guided mutation. A mutator takes seeds from the pool and applies graph specific mutations, including topology extension, predicate adjustment, clause insertion, path bound changes, and value mutation. \textsc{GRAF} executes the mutated Cypher queries on the target. An execution monitor records a feedback vector $\mathcal{F}=\langle t,|R|,s\rangle$, where $t$ is execution time, $|R|$ estimates result size, and $s$ denotes the execution status. This feedback guides subsequent mutations by suppressing slow or unproductive query shapes and promoting productive ones. \textsc{GRAF} retains queries that trigger abnormal engine behavior for reproduction, minimization, and root cause deduplication.

\subsection{Graph Context Aware Query Generation}
\label{sec:QueryGeneration}

\textsc{GRAF} separates abstract query structure from concrete graph data to produce valid and diverse queries. This section details the three steps of this generation process. First, graph context modeling extracts schema, topology, and value metadata from the target database. Second, skeleton generation utilizes a Large Language Model to synthesize a pool of abstract query structures. Finally, cascading dependency resolution binds these skeletons to the extracted context to generate executable queries.

\textit{Graph Context Modeling.}
\label{subsec:context}
Before generation begins, \textsc{GRAF} populates the target database with multiple labels, relationship types, properties, and topological patterns. The framework distills this environment into a formal context $\mathcal{C}=(\mathcal{M},\mathcal{G},\mathcal{V})$ to guide both instantiation and mutation.

\paragraph{Schema Metadata ($\mathcal{M}$)}
We define $\mathcal{M}=(L,R,P)$. $L$ represents the set of active node labels. $R\subseteq L\times T\times L$ captures the valid source label, relationship type, and destination label triples observed in the initialized graph. $P$ maps each entity to its property keys and data types. For instance, if $(\texttt{Customer},\texttt{ORDERED},\texttt{Product})\in R$, \textsc{GRAF} knows that after selecting \texttt{Customer} as a source, it can select \texttt{ORDERED} and bind the destination to \texttt{Product}. This mapping prevents the system from generating unavailable graph patterns or expressions with incompatible types.

\paragraph{Global Topology ($\mathcal{G}$)}
We define $\mathcal{G}=(|V|,|E|,\rho,\hat{d})$, where $|V|$ and $|E|$ denote the total nodes and relationships, $\rho$ represents density, and $\hat{d}$ estimates the graph diameter. \textsc{GRAF} utilizes these metrics to configure traversal bounds, \texttt{LIMIT} values, and other parameters controlling intermediate data sizes. While these statistics do not guarantee fast execution, they establish baselines for the target that improve upon hardcoded constants.

\paragraph{Type Specific Values ($\mathcal{V}$)}
For any given property $p$, $\mathcal{V}(p)$ provides a compatible typed value. It operates by either sampling existing graph data or synthesizing boundary and anomalous inputs. Observed values ensure that predicates match existing records. Meanwhile, boundary values such as numeric extremes, excessively long strings, and null pointers evaluate the storage and conversion logic of the database.

\setcounter{paragraph}{0}
\textit{Query Skeleton Generation via LLM.} 
Manually encoding every complex dependency of Cypher into a traditional grammar requires significant engineering effort. To address this, \textsc{GRAF} utilizes a Large Language Model~\cite{llm-fuzz} to synthesize a pool $\mathcal{T}$ of parameterized query skeletons focused on syntactic and structural diversity.
Figure~\ref{fig:prompt} details the generation prompt. It instructs the model to produce query templates native to the target dialect without filling concrete values. To ensure that the framework can accurately parse and instantiate these diverse skeletons, the prompt enforces three core structural rules:

\paragraph{Standardized Identifiers}
\textsc{GRAF} assigns distinct variable names based on entity roles, such as \texttt{n1} for nodes and \texttt{r1} for relationships. This convention makes variable roles and scope transitions explicit, which simplifies subsequent parsing, instantiation, and mutation.

\paragraph{Placeholder Identity}
Indexed placeholders encode equality constraints across the query skeleton. Repeated occurrences of \texttt{\$LABEL\_1} share the exact same binding. Conversely, \texttt{\$LABEL\_1} and \texttt{\$LABEL\_2} bind independently when the active graph context allows distinct choices.

\paragraph{Scope Annotations}
Expression placeholders explicitly define their legal scope. For instance, a standard predicate can only reference variables visible in the current clause, while a list comprehension predicate may also reference its captured iteration variable. This strict scoping prevents the framework from generating invalid variable references during instantiation.

\begin{figure}[t]
\centering
\begin{promptbox}
\footnotesize

\textbf{Task:} You are a fuzzing engineer generating parameterized query templates for testing the specific graph database system: [gdbms].

\vspace{0.5em}
\textbf{Requirements:}
\begin{enumerate}[leftmargin=1.5em]
    \item Output valid Cypher query skeletons.
    \item Do NOT fill concrete values.
    \item Standardize identifiers strictly to ensure the fuzzer can track variable scope:
    \begin{itemize}
        \item Nodes: \texttt{n1}, \texttt{n2}, \texttt{n3}, etc.
        \item Relationships: \texttt{r1}, \texttt{r2}, \texttt{r3}, etc.
        \item Paths: \texttt{p1}, \texttt{p2}, etc.
        \item Aliases or iteration variables: \texttt{v1}, \texttt{v2}, \texttt{v3}, etc.
    \end{itemize}

    \item Use the following base placeholders and append numeric suffixes (e.g., \texttt{\_1}, \texttt{\_2}) to represent distinct bindings:
    \begin{itemize}
        \item \texttt{\$LABEL\_i}: node label;
        \item \texttt{\$REL\_i}: relationship type;
        \item \texttt{\$N\_i}: numeric value or range bound;
        \item \texttt{\$DEPTH\_i}: path traversal depth;
        \item \texttt{\$PROP\_i}: property name;
        \item \texttt{\$VALUE\_i}: property value;
        \item \texttt{\$PREDICATE\_i}: Boolean condition over currently bound variables;
        \item \texttt{\$RETURN\_EXPR\_i}: return expression over currently bound variables;
        \item \texttt{\$PREDICATE\_FOR\_v}: condition scoped to iteration variable \texttt{v};
        \item \texttt{\$RETURN\_EXPR\_FOR\_v}: expression scoped to iteration variable \texttt{v}.
    \end{itemize}

    \item Maximize syntax and structure diversity to ensure maximal code coverage.
\end{enumerate}

\textbf{Output Requirement:} Generate 5000 query templates that are structurally complex, type-diverse, and completely native to the [gdbms] dialect. Output EXACTLY one template per line. 
\end{promptbox}
\caption{Prompt used for Cypher skeleton generation.}
\label{fig:prompt}
\end{figure}

After the model generates the initial candidates, \textsc{GRAF} applies four validation steps before a skeleton enters the final pool. First, placeholder validation confirms that all tags follow supported formats. Second, an ANTLR parser filters out syntactically invalid skeletons. Third, a scope analyzer checks variable definitions and their usage across clauses. Finally, a structural comparison removes duplicate skeletons.

\textit{Cascading Dependency Resolution.}
\label{subsec:instantiation}
When instantiating the generated skeletons, traditional fuzzers typically replace placeholders randomly. This approach frequently produces invalid queries that violate graph constraints, such as generating nonexistent topologies or accessing undefined properties for a given label.

\begin{figure*}[t]
    \centering
    \includegraphics[width=\linewidth]{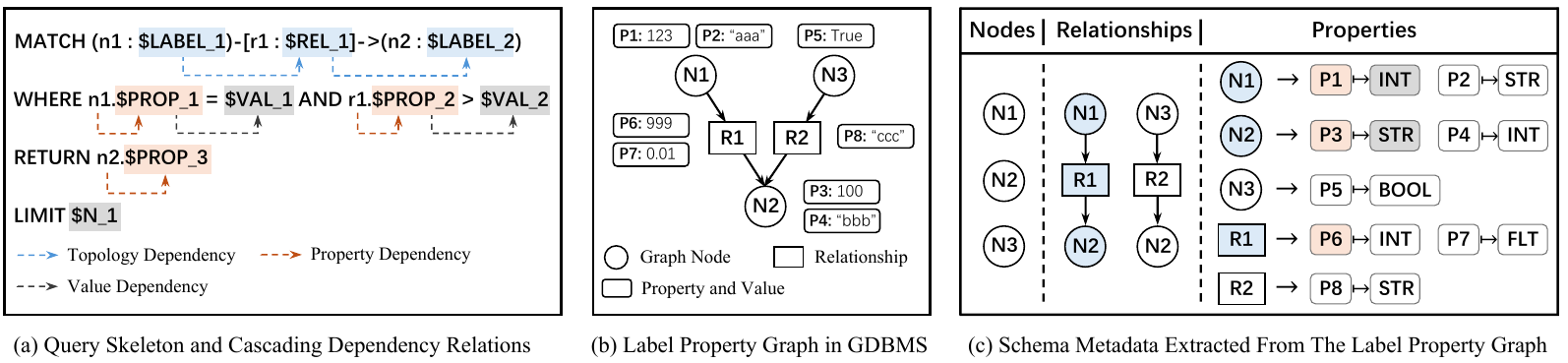}
    \caption{Cascading dependency resolution. (a) A parameterized skeleton where arrows indicate binding dependencies. (b) The graph initialized in the target. (c) Metadata extracted from the graph to constrain instantiation.}
    \label{fig:instantiation}
\end{figure*}

Instead of assigning values independently, \textsc{GRAF} models instantiation as a cascading dependency resolution process guided by the graph context $\mathcal{C}$. As Figure~\ref{fig:instantiation} illustrates, the framework follows a dependency chain with three stages. It first establishes the graph topology, then selects associated properties, and finally injects concrete data values. This order ensures that each refinement step remains within a valid semantic scope. Specifically, \textsc{GRAF} constructs the topological structure based on schema metadata $\mathcal{M} = (L, R, P)$. It samples an initial node label from $L$, and then selects valid relationships according to the directional constraints in $R$. These relationships subsequently determine the destination node labels, which ensures a valid graph pattern.

\begin{algorithm}[htbp]
\caption{Cypher Query Skeleton Instantiation}
\label{alg:instantiation}
\KwIn{Abstract query skeleton $T$, graph context $\mathcal{C}=(\mathcal{M},\mathcal{G},\mathcal{V})$}
\KwOut{Executable Cypher query $Q$}
$Q \gets T$\;

\tcp{\small\textbf{Phase 1: Topology Resolution}}
\ForEach{topological pattern $(n_{src},e_{rel},n_{dst})$ in $Q$}{
    $L_{src} \gets \text{ResolveLabel}(n_{src},\mathcal{M}_{L})$\;
    $R_{edge} \gets \text{Sample}(\text{GetValidOutgoing}(\mathcal{M}_{R},L_{src}))$\;
    $L_{dst} \gets \text{GetTargetLabel}(\mathcal{M}_{R},L_{src},R_{edge})$\;
    $Q \gets \text{UpdatePattern}(Q,L_{src},R_{edge},L_{dst})$\;
}

\tcp{\small\textbf{Phase 2: Property and Value Resolution}}
\ForEach{entity $E\in Q$ with property $p$ and value placeholder $v$}{
    $P_{meta} \gets \text{Sample}(\mathcal{M}_{P}(E.\text{label}))$\;
    $Val \gets \mathcal{V}.\text{Generate}(P_{meta}.\text{type})$\;
    $Q \gets \text{Bind}(Q,p\to P_{meta}.\text{name},v\to Val)$\;
}

\tcp{\small\textbf{Phase 3: Global Parameter Alignment}}
$Q \gets \text{AlignGlobalBounds}(Q,\mathcal{G})$\;
\Return $Q$\;
\end{algorithm}

To illustrate Algorithm~\ref{alg:instantiation} using the example in Figure~\ref{fig:instantiation}, consider transforming an abstract skeleton $T$ into an executable query $Q$ using the graph context $\mathcal{C}$. 
In Phase 1, given a topological pattern $(n_{src}, e_{rel}, n_{dst})$ such as \texttt{MATCH (n1:\$LABEL\_1)-[r1:\$REL\_1]->\allowbreak{}(n2:\$LABEL\_2)}, \textsc{GRAF} resolves the source label $L_{src}$ by sampling \texttt{N1} from the schema $\mathcal{M}_L$. This choice restricts the valid outgoing edge $R_{edge}$ to \texttt{R1}, which then determines the target label $L_{dst}$ as \texttt{N2}. This cascading dependency ensures that the resulting graph patterns comply with the schema. 
In Phase 2, for each entity $E$ containing property ($p$) and value ($v$) placeholders, \textsc{GRAF} samples valid property metadata $P_{meta}$ from $\mathcal{M}_P$. For instance, the framework instantiates $p$ for \texttt{N1} as \texttt{P1(INT)}. Subsequently, the value constructor $\mathcal{V}$ synthesizes a concrete value $Val$ of the matching type. Thus, \textsc{GRAF} can assign \texttt{123} through data sampling or \texttt{2147483647} through boundary generation to $v$. 
Finally, in Phase 3, \textsc{GRAF} aligns global parameters with the global topology $\mathcal{G}$. Rather than using fixed bounds, it limits parameters like \texttt{\$N\_1} in a \texttt{LIMIT} clause proportionally to the total vertex count $|\mathcal{G}.V|$. By integrating all components of $\mathcal{C}$, this instantiation process ensures that $Q$ remains topologically and semantically valid.

\subsection{Execution State Guided Query Mutation}
\label{sec:feedback_mutation}

To explore execution behaviors beyond the initial seeds, \textsc{GRAF} introduces a query mutation framework driven entirely by execution state feedback. This section details the two components of this process. First, it defines five graph specific mutation operators that modify the query syntax tree while maintaining structural and contextual validity. Second, it introduces a feedback guided exploration strategy. By tracking execution time, result size, and system status, \textsc{GRAF} dynamically adapts these operators to avoid unproductive queries and steer exploration toward deep engine logic.

\textit{Graph Specific Mutation Operators.}
\textsc{GRAF} targets the primary factors that influence graph query execution: topology, traversal depth, predicate selectivity, clause pipelines, and data values. To explore these dimensions, the framework applies five corresponding families of mutation operators.

\paragraph{Topology Mutation}
This operator dynamically inserts, removes, or reconnects patterns within \texttt{MATCH} clauses. When extending a bound node, \textsc{GRAF} samples relationship triples compatible with the node label. This transformation prompts the database optimizer to recalculate join orders and pattern planning without breaking query validity.

\paragraph{Depth and Volume Mutation}
This operator modulates path bounds, \texttt{LIMIT} configurations, and variables controlling intermediate data sizes. Initial values derive from $\mathcal{G}$ and adjust continuously based on execution feedback. \textsc{GRAF} scales these parameters gradually, preventing the framework from stalling on variants that time out.

\paragraph{Predicate Mutation}
This operator manipulates \texttt{WHERE} expressions through insertion, replacement, deletion, or relaxation. Property references rely on $\mathcal{M}.P$, while literals draw from $\mathcal{V}$. By altering predicate selectivity while maintaining type correctness, this operator evaluates index selection and downstream filtering routines.

\paragraph{Clause Insertion}
This operator injects pipeline components such as \texttt{WITH}, aggregations, and supported subqueries into legal tree positions. Inputs bind dynamically to variables visible at the insertion point. This evaluates intermediate result scheduling and variable propagation across clauses.

\paragraph{Value Mutation}
This operator substitutes a literal with an observed, boundary, or anomalous value of a compatible type. Observed values increase the probability of database hits, whereas boundary values trigger edge cases in the conversion and storage modules.

\setcounter{paragraph}{0}
\textit{Feedback Guided Exploration.}
Random mutation in graph queries often leads to two inefficiencies. First, valid queries frequently return empty result sets. Second, complex queries stall in timeouts. To distinguish these states, \textsc{GRAF} records a dynamic feedback vector $\mathcal{F}=\langle t,|R|,s\rangle$, where $t$ represents execution time, $|R|$ denotes result size, and $s$ characterizes the termination state.

\paragraph{Fast Queries with Empty Results}
When a query terminates quickly with $|R|=0$, \textsc{GRAF} treats the query as overly constrained. It subsequently increases the probability of predicate relaxation, topology adjustment, and replacement with observed values. These adjustments adapt the underlying structure to ensure that subsequent projections and aggregations receive materialized data.

\paragraph{High Cost and Timed Out Queries}
While slow queries often expose logic flaws, increasing traversal depth on queries that already time out provides little benefit. For the descendants of these expensive seeds, \textsc{GRAF} penalizes operators that increase depth or volume. Instead, it favors bounded reductions or mutations to other query components to maintain efficient exploration. 

\paragraph{Abnormal Terminations}
Executions resulting in crashes, assertion failures, sanitizer alerts, or reproducible resource exhaustion mark a query for priority analysis. \textsc{GRAF} retains these inputs and applies localized mutations to isolate the clauses responsible for the failure. Final validation requires clean state reproduction and root cause deduplication rather than directly counting raw abnormal executions as bugs.

Execution time and result size serve as active guidance signals to allocate mutation effort. Code coverage remains a separate evaluation metric, while this state feedback allows \textsc{GRAF} to sustain continuous exploration within the graph query space.


\section{Implementation}
\label{sec:implementation}

We implemented \textsc{GRAF} in approximately 15K lines of Python code. The system comprises a skeleton generator and filter, a Cypher frontend built on ANTLR~\cite{antlr}, a graph context collector, a context guided instantiator, a syntax tree mutator, an execution monitor, and individual adapters for Memgraph, Neo4j, FalkorDB, KuzuDB, NebulaGraph, and RedisGraph. Memgraph and Neo4j are accessed through the Bolt protocol, FalkorDB and RedisGraph through Redis graph commands, KuzuDB through its embedded API, and NebulaGraph through its native client. 

To ensure structural diversity while minimizing overhead, we synthesize 5,000 query skeletons. This volume provides sufficient syntactic complexity because our cascading dependency resolution dynamically instantiates these skeletons into a vast combinatorial space of executable queries. Furthermore, leveraging a Large Language Model (Gemini-3.1-pro) for this one-time synthesis incurs a negligible cost of less than \$10. This highly cost effective approach eliminates the substantial manual engineering effort traditionally required to develop and maintain complex grammar generators.

During fuzzing, \textsc{GRAF} loads the filtered skeleton pool, refreshes the graph context when the target state changes, instantiates and reparses each candidate, applies graph specific mutations, executes the resulting query, and updates mutation selection using execution state feedback. Each adapter manages graph initialization, query submission, timeout enforcement, health checking, failure collection, and recovery. The monitor records process termination, assertion output, sanitizer diagnostics when available, timeouts, memory limit events, and connection failures.
The core fuzzing process operates entirely as a black box and does not use source coverage to select inputs. For evaluation purposes only, C and C++ targets are built with LLVM~\cite{llvm} source coverage, while Neo4j is instrumented with JaCoCo~\cite{jacoco}. Coverage reports are generated from periodically collected profiles.


\section{Evaluation}
\label{sec}

We evaluate \textsc{GRAF} from four perspectives: its ability to
exercise target code, generate valid queries, benefit from its major
design components, and expose previously unknown failures in
production GDBMSs. Specifically, our evaluation addresses the
following research questions:

\begin{itemize}[leftmargin=*, labelindent=0pt]

\item \textbf{RQ1 (Code Coverage):}
How much target code does \textsc{GRAF} cover compared with the baselines over 12 hours?

\item \textbf{RQ2 (Query Validity):}
How effectively does \textsc{GRAF} generate syntactically and semantically valid graph queries?

\item \textbf{RQ3 (Ablation Study):}
How does each major component of \textsc{GRAF} contribute to
query validity and code coverage?

\item \textbf{RQ4 (Bug Detection):}
What previously unknown bugs can
\textsc{GRAF} discover in popular GDBMSs?
\end{itemize}

\subsection{Experimental Setup}
\label{sec:setup}

\paragraph{Tested GDBMSs}
We evaluate \textsc{GRAF} on six mainstream GDBMSs that support Cypher or related graph query languages: Neo4j, Memgraph, RedisGraph, NebulaGraph, FalkorDB, and KuzuDB. Table~\ref{tab:gdbms_info} summarizes their representatively tested versions, popularity, rankings~\cite{dbengines-ranking} and lines of code (LoC). Each experiment starts from a clean database initialized by a dedicated graph initializer.

\begin{table}[htbp]
\centering
\caption{Evaluated GDBMSs.}
\label{tab:gdbms_info}
\begin{tabular}{l c c c c}
\toprule
\textbf{GDBMS} & \textbf{Version} & \textbf{GitHub Stars} & \textbf{Ranking} & \textbf{LoC} \\
\midrule
Neo4j       & 5.26.27 & 16.8k & 1  & 1.89M \\
Memgraph    & 3.7.1   & 4.2k  & 10 & 498K  \\
RedisGraph  & 2.12.10 & 2k    & 4  & 1.30M \\
NebulaGraph & 3.8.0   & 12.2k & 9  & 344K  \\
FalkorDB    & 4.16.1  & 4.7k  & 25 & 1.25M \\
KuzuDB      & 0.11.3  & 3.8k  & 37 & 14.2M \\
\bottomrule
\end{tabular}
\end{table}

\paragraph{Baselines}
We compare \textsc{GRAF} with Dinkel~\cite{Dinkel}, BUZZBEE~\cite{BUZZBEE}, and AFL++~\cite{aflpp}. Dinkel represents state aware graph query generation, BUZZBEE represents structure aware mutation, and AFL++ represents generic coverage guided byte mutation. We use their official releases and retain their original query generation, mutation, and scheduling policies. If a baseline lacks native support for a target, we adapt only the database connection, query transport, and result handling interfaces without modifying the generation logic.

\paragraph{Experimental Infrastructure}
We conduct experiments on a 64 bit Ubuntu server with 72 logical CPU cores and 256~GB of memory. Each fuzzer and target combination runs in an isolated Docker container limited to four CPU cores and 16~GB of memory. Each coverage experiment lasts 12 hours. We repeat all configurations using independent random seeds, and the plots report the median values with 95\% confidence intervals.

\begin{figure}[t]
    \centering
    \includegraphics[width=\linewidth]{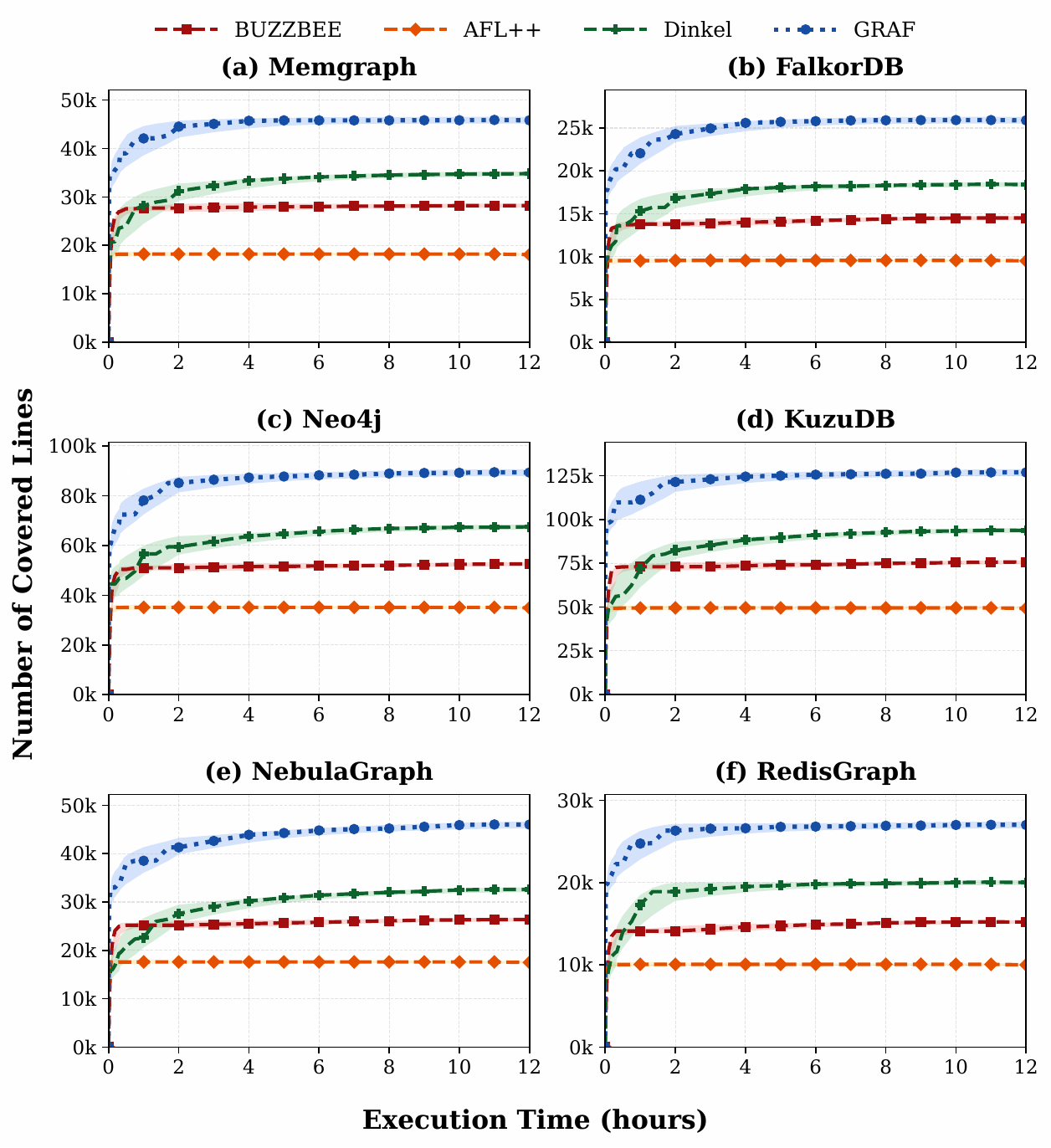}
    \caption{Coverage growth over 12 hours. The curves report the median number of covered code lines, and shaded regions denote 95\% confidence intervals.}
    \label{fig:coverage}
\end{figure}

\subsection{RQ1: Code Coverage}
\label{sec:rq1}

Figure~\ref{fig:coverage} reports line coverage growth over 12 hours. \textsc{GRAF} achieves the highest final coverage on every target: 45.8K lines on Memgraph, 25.9K on FalkorDB, 89.3K on Neo4j, 126.9K on KuzuDB, 46.0K on NebulaGraph, and 27.1K on RedisGraph. Relative to the strongest baseline for each target, \textsc{GRAF} improves coverage by 31.6\% to 41.1\%. Compared individually, it covers 31.6\% to 77.3\% more lines than Dinkel, 62.4\% to 79.8\% more than BUZZBEE, and 150.6\% to 172.2\% more than AFL++.

The coverage curves exhibit distinct growth patterns. AFL++ grows quickly initially because small mutations of seed queries easily reach the parsing and basic validation code. However, its growth stalls as byte level mutations increasingly corrupt the Cypher syntax. Dinkel maintains a high acceptance rate, but its generated queries lack structural variety. BUZZBEE explores more syntactic variants but operates without runtime topology and value information. In contrast, \textsc{GRAF} sustains continuous coverage growth. The skeleton pool introduces diverse pipeline structures, and context guided instantiation ensures that these complex queries remain executable.

\begin{figure}[t]
    \centering
    \includegraphics[width=\linewidth]{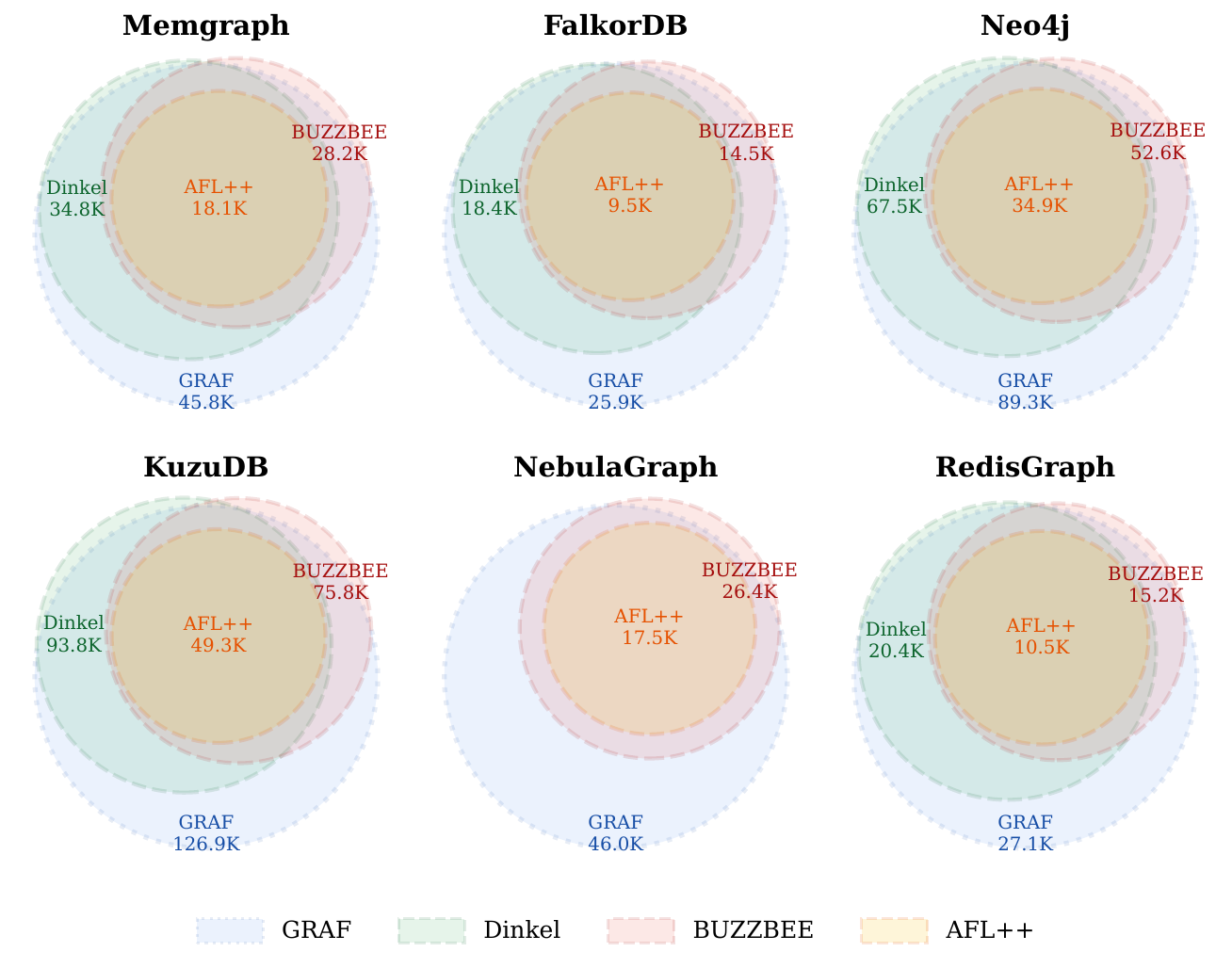}
    \caption{Line coverage overlap at 12 hours. \textsc{GRAF} covers the baseline intersection and achieves the highest unique coverage.}
    \label{fig:coverage_overlap}
\end{figure}

Figure~\ref{fig:coverage_overlap} compares the sets of covered lines at the end of the 12 hour experiment. The baselines share a substantial intersection of covered lines, which primarily correspond to the frontend parser, basic semantic checks, and shallow execution paths. The coverage of AFL++ is almost entirely contained within the regions reached by the structured generators (Dinkel and BUZZBEE). \textsc{GRAF} successfully covers this common baseline intersection. More importantly, \textsc{GRAF} contributes the largest set of uniquely covered lines on every target. This indicates that the coverage improvement stems from reaching deep implementation regions, rather than merely executing more accepted queries.

\vspace{8pt}
\begin{rqanswer}
\textbf{Answer to RQ1:} \textsc{GRAF} consistently achieves the highest code coverage across all six targets, improving upon the strongest baseline by 31.6\% to 41.1\%.
\end{rqanswer}\subsection{RQ2: Query Validity}
\label{sec:rq2}

A query is \emph{valid} if the target accepts it past syntax and semantic validation and begins execution. Normal completion, timeout, and confirmed target failure are therefore valid outcomes; parser and semantic analysis rejections are invalid. For each target, we run each fuzzer for one hour and compute
\[
\mathrm{VQR}=\frac{\#\text{queries accepted for execution}}{\#\text{queries submitted}}\times 100\%.
\]
This metric measures whether generated queries pass frontend checks and reach the execution engine. It does not by itself measure code coverage or bug finding ability.

\begin{table}[t]
\centering
\caption{Valid query rate across target GDBMSs.}
\label{tab:query_validity}
\begin{tabular}{lcccc}
\toprule
\textbf{GDBMS} & \textbf{\textsc{GRAF}} & \textbf{Dinkel} & \textbf{BUZZBEE} & \textbf{AFL++} \\
\midrule
Memgraph    & 66.4\% & 81.2\% & 54.8\% & 2.2\% \\
FalkorDB    & 75.8\% & 83.5\% & 61.7\% & 2.0\% \\
Neo4j       & 76.7\% & 84.1\% & 63.9\% & 2.2\% \\
KuzuDB      & 72.6\% & 78.4\% & 58.2\% & 2.1\% \\
NebulaGraph & 61.8\% &  --    & 49.6\% & 1.9\% \\
RedisGraph  & 74.3\% & 82.7\% & 60.4\% & 2.3\% \\
\bottomrule
\end{tabular}
\end{table}

Table~\ref{tab:query_validity} shows that \textsc{GRAF}'s valid query rate ranges from 61.8\% to 76.7\%. Dinkel achieves a higher rate on supported targets because it uses more conservative templates. BUZZBEE generates more invalid queries because it does not use runtime graph information to match labels, relationship types, properties, and values. AFL++ accepts only 1.9\% to 2.3\% of submissions because byte level mutations often corrupt Cypher tokens and grammar.

These results show that most \textsc{GRAF} queries pass frontend checks even though its skeletons include varied clause combinations, traversals, expressions, and dependencies across clauses. Its valid query rate is lower than Dinkel's, but it remains consistently higher than those of BUZZBEE and AFL++ on every supported target.

\vspace{8pt}
\begin{rqanswer}
\textbf{Answer to RQ2:} \textsc{GRAF} maintains a valid query rate of 61.8\% to 76.7\%. Context guided instantiation balances query validity with structural complexity, avoiding the limited search space of strictly conservative generators.
\end{rqanswer}

\begin{table*}[t]
\centering
\caption{Ablation study: average valid query rate and line coverage across targets. Parenthesized values report relative degradation from the complete system.}
\label{tab:ablation_coverage}
\renewcommand{\arraystretch}{1.15}

\newcommand{\mr}[2]{\makebox[3.2em][r]{#1}\hspace{0.2em}\makebox[3.2em][l]{#2}}
\newcommand{\mb}[1]{\makebox[3.2em][r]{#1}\hspace{0.2em}\makebox[3.2em][l]{}}

\resizebox{\textwidth}{!}{
\begin{tabular}{l c c c c c c c}
\toprule

\multirow{2}{*}[-3pt]{\textbf{Fuzzer Variant}} 
& \multirow{2}{*}[-3pt]{\textbf{Valid Query Rate}} 
& \multicolumn{6}{c}{\textbf{Line Coverage}} \\
\cmidrule(l){3-8}
& & \textbf{Memgraph} & \textbf{FalkorDB} & \textbf{Neo4j} 
& \textbf{KuzuDB} & \textbf{NebulaGraph} & \textbf{RedisGraph} \\
\midrule

\rowcolor[gray]{0.92}
\textsc{GRAF} (Full)
& \mb{71.3\%}
& \mb{45.8K}  & \mb{25.9K}  & \mb{89.3K}  
& \mb{126.9K} & \mb{46.0K}  & \mb{27.1K}  \\

w/o Graph Initializer
& \mr{65.6\%}{\textit{(-8\%)}}
& \mr{23.4K}{\textit{(-49\%)}} & \mr{12.4K}{\textit{(-52\%)}} & \mr{46.4K}{\textit{(-48\%)}}
& \mr{57.1K}{\textit{(-55\%)}} & \mr{25.3K}{\textit{(-45\%)}} & \mr{12.4K}{\textit{(-54\%)}} \\

w/o Skeleton Generator
& \mr{29.9\%}{\textit{(-58\%)}}
& \mr{12.8K}{\textit{(-72\%)}} & \mr{8.0K}{\textit{(-69\%)}} & \mr{26.8K}{\textit{(-70\%)}}
& \mr{31.7K}{\textit{(-75\%)}} & \mr{12.0K}{\textit{(-74\%)}} & \mr{7.8K}{\textit{(-71\%)}} \\

w/o Skeleton Instantiator
& \mr{10.7\%}{\textit{(-85\%)}}
& \mr{7.3K}{\textit{(-84\%)}} & \mr{4.9K}{\textit{(-81\%)}} & \mr{15.2K}{\textit{(-83\%)}}
& \mr{17.8K}{\textit{(-86\%)}} & \mr{6.9K}{\textit{(-85\%)}} & \mr{4.9K}{\textit{(-82\%)}} \\

w/o Mutator
& \mr{69.1\%}{\textit{(-3\%)}}
& \mr{27.9K}{\textit{(-39\%)}} & \mr{16.8K}{\textit{(-35\%)}} & \mr{55.4K}{\textit{(-38\%)}}
& \mr{76.1K}{\textit{(-40\%)}} & \mr{26.7K}{\textit{(-42\%)}} & \mr{17.3K}{\textit{(-36\%)}} \\

w/o Execution Monitor
& \mr{67.7\%}{\textit{(-5\%)}}
& \mr{33.4K}{\textit{(-27\%)}} & \mr{18.4K}{\textit{(-29\%)}} & \mr{64.3K}{\textit{(-28\%)}}
& \mr{93.9K}{\textit{(-26\%)}} & \mr{35.0K}{\textit{(-24\%)}} & \mr{18.6K}{\textit{(-31\%)}} \\

\bottomrule
\end{tabular}
}
\end{table*}

\subsection{RQ3: Component Contribution}
\label{sec:rq3}

Table~\ref{tab:ablation_coverage} disables one component at a time and reports average VQR and per-target line coverage. The full system attains a 71.3\% average VQR and the highest coverage on all six systems.

\paragraph{Semantic instantiation}
Removing the Skeleton Instantiator causes the largest loss: VQR falls from 71.3\% to 10.7\% (an 85\% relative decrease), and coverage drops by 81\% to 86\%. Independent substitution therefore fails to preserve the topology, type, and scope dependencies needed to pass frontend validation. Removing the Skeleton Generator reduces VQR by 58\% and coverage by 69\% to 75\%, confirming that semantic instantiation alone cannot compensate for a limited structural seed space.

\paragraph{State construction and mutation}
Without the Graph Initializer, coverage drops by 45\% to 55\%. A nontrivial graph state is thus necessary even when statements remain syntactically valid. Removing the Mutator changes VQR by only 3\% but decreases coverage by 35\% to 42\%, indicating that mutation primarily expands the explored execution space rather than merely repairing invalid inputs.

\begin{figure}[t]
    \centering
    \includegraphics[width=\linewidth]{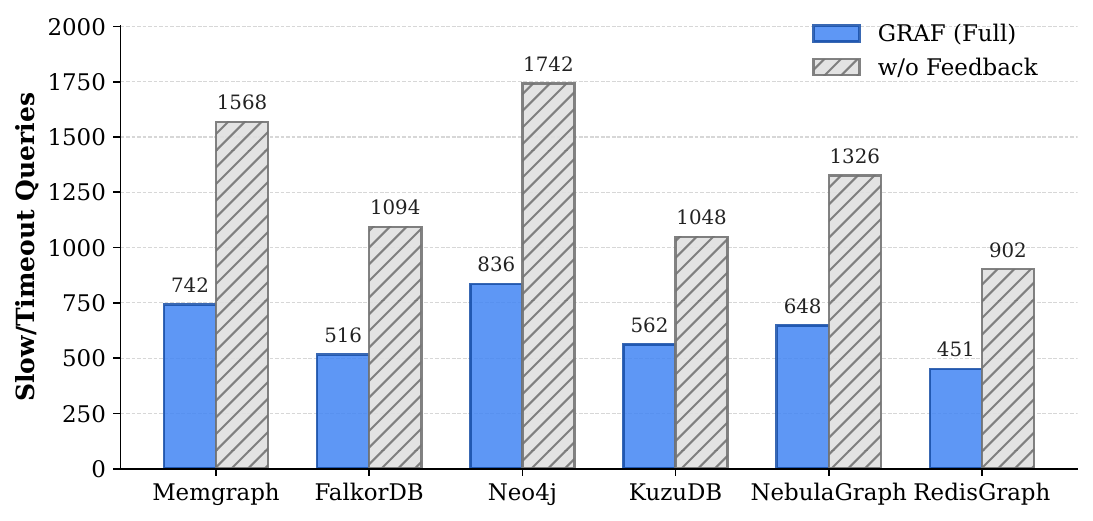}
    \caption{Slow executions with and without execution state feedback.}
    \label{fig:slow_queries}
\end{figure}

\paragraph{Execution state feedback}
Removing the Execution Monitor reduces VQR by 5\% but coverage by 24\% to 31\%. Figure~\ref{fig:slow_queries} explains the difference: without feedback, the number of slow or timed out queries (exceeding 60 seconds) increases from 3,755 to 7,680. Across individual targets, the complete system reduces such executions by 46.4\% to 52.8\%. The feedback mechanism therefore avoids repeatedly increasing the depth or intermediate result size of query variants that are already too expensive, allowing more valid queries to be tested.

\vspace{8pt}
\begin{rqanswer}
\textbf{Answer to RQ3:} Every component of \textsc{GRAF} is indispensable. The instantiator is the primary driver of query validity, while the remaining modules ensure sustained coverage growth and deep state exploration.
\end{rqanswer}

\subsection{RQ4: Bug Detection}
\label{sec:rq4}

We consider failures that compromise memory safety or system availability: segmentation faults, use after free errors, stack overflows, assertion failures, undefined behavior, and out of memory (OOM) failures. Semantic logic errors (e.g., incorrect result sets) are excluded, as our oracle evaluates system stability. We deduplicate raw crashes by root cause, and Table~\ref{tab:severe_bugs} reports only those bugs that developers have confirmed.

\begin{table}[t]
\centering
\caption{Confirmed bugs detected by \textsc{GRAF}.}
\label{tab:severe_bugs}
\resizebox{\columnwidth}{!}{%
\begin{tabular}{lll}
\toprule
\textbf{Project (\#Bugs)} & \textbf{Component} & \textbf{Bug Type} \\
\midrule
\multirow{6}{*}{Memgraph (9)}
& Pattern Matcher & Segmentation fault (2) \\
& Projection Operator & Use after free (2) \\
& Built in Function & Segmentation fault (1) \\
& Built in Function & Undefined Behavior (1) \\
& Path Expansion Operator & Segmentation fault (1) \\
& Transaction Manager & Out of memory (2) \\
\midrule
\multirow{9}{*}{FalkorDB (11)}
& Query Planner & Assertion failure (1) \\
& Expression Evaluator & Out of memory (1) \\
& Write/Update Operator & Segmentation fault (2) \\
& Edge Create Operator & Segmentation fault (1) \\
& Path Expansion Operator & Segmentation fault (1) \\
& List Comprehension Operator & Out of memory (1) \\
& Aggregation Operator & Out of memory (1) \\
& Property Accessor & Use after free (2) \\
& Array Allocator & Out of memory (1) \\
\midrule
Neo4j (2)
& Query Runtime & Java heap space OOM (1) \\
& Query Runtime & Undefined Behavior (1) \\
\midrule
\multirow{2}{*}{KuzuDB (2)}
& Query Parser & Stack overflow (1) \\
& Built in Function & Out of memory (1) \\
\midrule
\multirow{2}{*}{NebulaGraph (2)}
& Expression Evaluator & Out of memory (1) \\
& Path Expansion Operator & Segmentation fault (1) \\
\midrule
\multirow{5}{*}{RedisGraph (6)}
& Query Parser & Stack overflow (1) \\
& Property Accessor & Use after free (1) \\
& List Comprehension Operator & Out of memory (1) \\
& Expression Evaluator & Out of memory (1) \\
& Write/Update Operator & Segmentation fault (2) \\
\bottomrule
\end{tabular}%
}
\end{table}

Across all evaluation campaigns, \textsc{GRAF} identified 34 previously unknown bugs. Table~\ref{tab:severe_bugs} summarizes the 32 confirmed bugs across the six evaluated systems, with the majority being OOM failures (11) and segmentation faults (11). At the time of writing, 23 of these bugs have been assigned CVE identifiers. These bugs span various core components, including pattern matching, projection, path expansion, aggregation, property access, and transaction management. This broad distribution demonstrates that \textsc{GRAF}'s structurally complex queries successfully bypass frontend validation to evaluate backend operators.

\begin{table}[t]
  \centering
  \caption{Unique bugs triggered within 12 hours.}
  \label{tab:unique_bugs}
  \begin{tabular}{lcccc}
    \toprule
    \textbf{Target GDBMS} & \textbf{Dinkel} & \textbf{BUZZBEE} & \textbf{AFL++} & \textbf{\textsc{GRAF}} \\
    \midrule
    Memgraph    & 1 & 1 & 0 & \textbf{8} \\
    FalkorDB    & 2 & 1 & 0 & \textbf{7} \\
    Neo4j       & 0 & 0 & 0 & \textbf{2} \\
    KuzuDB      & 0 & 0 & 0 & \textbf{2} \\
    NebulaGraph & -- & 0 & 0 & \textbf{2} \\
    RedisGraph  & 1 & 0 & 0 & \textbf{4} \\
    \midrule
    \textbf{Total} & \textbf{4} & \textbf{2} & \textbf{0} & \textbf{25} \\
    \bottomrule
  \end{tabular}
\end{table}

To provide a standardized comparison, Table~\ref{tab:unique_bugs} reports the unique bugs triggered during a strictly controlled 12 hour evaluation. Under identical resource constraints, \textsc{GRAF} triggered 25 unique bugs. In contrast, Dinkel, BUZZBEE, and AFL++ triggered only four, two, and zero bugs, respectively. \textsc{GRAF} uncovered more bugs than all three baselines combined.

These empirical results align directly with our coverage and validity measurements. Dinkel's conservative grammar frequently generates valid statements but fails to synthesize the compound execution patterns required to evaluate the backend. BUZZBEE explores a wider syntactic space but lacks the runtime context necessary for semantic validity. AFL++ mutates byte streams, rarely preserving the Cypher semantics required to pass the parser. Conversely, \textsc{GRAF} successfully generates queries that intertwine multi hop traversals, optional matching, list comprehensions, and sequential update pipelines, systematically exposing the backend components where these failures reside.

\vspace{8pt}
\begin{rqanswer}
\textbf{Answer to RQ4:} \textsc{GRAF} discovered 34 previously unknown bugs across six targets, yielding 32 confirmations and 23 CVEs. Under controlled evaluation, it triggered 25 unique bugs, uncovering four times as many bugs as all baselines combined.
\end{rqanswer}

\section{RELATED WORK}
In this section, we focus on the relevant work on GDBMS differential test, metamorphic testing and DBMS fuzzing. 

\subsection{Differential Testing for GDBMSs}

Differential testing compares the outputs of the same query across multiple GDBMSs to identify inconsistencies that reveal potential bugs. GDsmith~\cite{GDsmith} systematically generates random graph schemas and Cypher queries to identify inconsistencies among Cypher-based GDBMSs. For imperative graph querying languages, Grand~\cite{Grand} introduces a randomized differential testing approach tailored for Gremlin-based systems, generating valid Gremlin traversals to validate query execution correctness across multiple implementations. Furthermore, GQS~\cite{GQS} focuses on the synthesis of graph queries, ensuring both syntactic validity and semantic structural constraints, thereby generating high-quality test cases for comprehensive differential testing across diverse graph engines. Different from these multi-engine approaches, our work focuses on deep bug discovery within a single GDBMS through feedback-guided mutation generation fuzzing.

\subsection{Metamorphic Testing for GDBMSs}

Metamorphic testing validates a GDBMS by checking whether semantically equivalent query variants produce consistent results under defined transformations~\cite{cypherequal}. GRev~\cite{GRev} utilizes Equivalent Query Rewriting (EQR) rules to generate semantically equivalent graph queries, checking result consistencies. To construct more robust metamorphic relations, GraphGenie~\cite{GraphGenie} introduces injective and surjective graph query transformations, which systematically alter the structural constraints and cardinality mappings of the original queries to uncover subtle logic bugs. GDBMeter~\cite{GDBMeter} adapts the Ternary Query Partitioning (TQP) concept to graph databases, partitioning a core graph query into multiple sub-queries whose aggregated results should logically reconstruct the original result set. Similarly, QuDi~\cite{QuDi} addresses the unique challenges of imperative queries by disassembling complex Gremlin traversals into multiple independent operational steps, verifying the consistency between the original query and the aggregated disassembled results.  Gamera~\cite{Gamera} proposes graph-aware metamorphic relations, which directly leverage inherent graph topological properties, such as path connectivity and structural equivalence, to define novel test oracles without relying on query semantics alone. Additionally, Dinkel~\cite{Dinkel} applies fine-grained clause- and expression-level semantics-preserving transformations to a Cypher query and checks whether the transformed query preserves the original result. Different from these metamorphic tools, our work relies on execution feedback rather than pre-defined equivalence rules to steer query generation.

\subsection{Fuzzing for DBMSs}

Existing DBMS fuzzing techniques~\cite{UNICORN,arg,EDC,SRS,Hulk,ConstantOD,DQP,TLP,Norec,PQS,SQUIRREL,sqlsmith2018,llmrewriter,Reusefuzz,QPG} include generation-based and mutation-based approaches. Generation-based fuzzers synthesize test cases using predefined rules or constraint solvers (e.g., SQLsmith~\cite{sqlsmith2018}). However, the NP-complete ~\cite{SQUIRREL} nature of query generation often forces restricting structural diversity to maintain valid database states. To systematically detect logic bugs, SQLancer~\cite{TLP,PQS,Norec} automatically synthesizes queries and validates their correctness using advanced logical oracles like Ternary Logic Partitioning (TLP~\cite{TLP}). On the other hand, to achieve deeper code exploration, mutation-based fuzzers leverage coverage feedback. ARG~\cite{arg} proposes an abstract rule-based testing method to test the correctness and robustness of query rewriters through intelligent generation and built-in SQL queries. While traditional mutational tools (e.g., AFL~\cite{afl2017}) struggle to pass the strict syntax parsers of complex database engines, recent advancements have significantly improved input validity. Specifically, SQUIRREL~\cite{SQUIRREL} designs an intermediate representation to enable syntax-preserving and semantics-guided mutations. RATEL further adapts mutational fuzzing for enterprise-level DBMSs by enhancing feedback precision and providing online root-cause investigation. Furthermore, UNICORN~\cite{UNICORN} targets time-series databases by combining syntax-preserved and time-series-guided mutations to generate valid queries. Lastly, GRIFFIN~\cite{Griffin} employs a metadata graph to mutate test cases efficiently in a completely grammar-free manner. BUZZBEE~\cite{BUZZBEE} targets cross-DBMS testing by abstracting database-independent input generation, mutation, and feedback mechanisms, enabling fuzzing to be ported to different database systems with low adaptation cost. Different from DBMS fuzzers, our framework handles complex graph topologies and cascading dependencies, specifically targeting structural bugs in GDBMSs.

\section{Conclusion}
\label{sec:conclusion}

This paper presented \textsc{GRAF}, a black box fuzzing framework for graph database management systems. Testing these systems is challenging because valid queries require strict adherence to syntax and underlying graph topology. \textsc{GRAF} addresses this by separating abstract query structure from concrete data bindings using a Large Language Model and cascading dependency resolution. Furthermore, it employs execution state guided mutation, utilizing execution time, result size, and system status to steer exploration toward deep engine logic without relying on traditional code coverage. We evaluated \textsc{GRAF} on six widely used graph databases against three recent fuzzers. Under identical constraints, \textsc{GRAF} consistently achieved the highest code coverage, improving upon the strongest baselines by 31.6\% to 41.1\%. This exploration led to the discovery of 34 previously unknown bugs, yielding 32 developer confirmations and 23 CVE identifiers. These results demonstrate that decoupling query structure from runtime context, combined with external execution feedback, effectively tests complex database engines. Future work will extend these context guided techniques to uncover semantic logic bugs in other specialized database paradigms.


\balance
\bibliographystyle{unsrt}
\bibliography{references.bib}

\end{document}